\documentclass[twocolumn,showpacs,preprintnumbers,amsmath,amssymb,prl]{revtex4}
\usepackage{graphicx}
 \newcommand{\bea}{\begin{eqnarray}}
 \newcommand{\eea}{\end{eqnarray}}
 
\begin{document} 
\draft
\title{Superfluid equation of state of dilute composite bosons}
\author{X. Leyronas and R. Combescot}
\affiliation{ Laboratoire de Physique Statistique, Ecole Normale Sup\'erieure, 24 rue Lhomond, 75231 Paris Cedex 05, France}
\begin
{abstract}
We present an exact theory of the BEC-BCS crossover in the BEC regime, which treats explicitely
dimers as made of two fermions. We apply our framework, at zero temperature,
to the calculation of the equation of state. We find that, when expanding the chemical potential in powers of the density $n$ up to the Lee-Huang-Yang order, proportional to $n^{3/2}$, the result
is identical to the one of elementary bosons in terms of the dimer-dimer scattering length $a_M$, 
the composite nature of the dimers appearing  only in the next order term proportional to $n^2$.
\end{abstract}
\maketitle

The BEC-BCS crossover first considered by Leggett \cite{leggett}, and the recent experimental realization of Bose Einstein Condensates (BEC) of molecules made of
fermionic atoms \cite{Grei,Joch,Zwie,Bour} have motivated a number of theoretical works. Indeed, thanks to Feshbach resonances, it is experimentally possible, with two fermions of mass $m$ ($^6$Li or $^{40}$K) in different hyperfine states (we denote them as 'spin' $\uparrow $ and $\downarrow $), with scattering length $a$, to realize weakly bound molecules, or dimers, with binding energy $E_b=1/ma^2$ (we take $\hbar =1$ in the paper). In particular one may obtain a dilute condensate of molecules. A crucial quantity controling the physics of the condensate in this regime is the dimer-dimer scattering length $a_M$. 
This is however a highly non-trivial quantity to calculate, since one has to solve a four-body problem to find it.
In the case of a broad resonance, one finds $a_M=0.6\, a$ by solving the Schr\"odinger equation \cite{PSS} or resumming the diagrammatic series \cite{condmat4par}.
The study of a Bose-Einstein condensate of composite bosons, where all the theory is formulated in terms of fermions only, was started a long time ago \cite{popov,kk} with steady progress. Quite recently, Pieri and Strinati \cite{pieristrinati} derived the Gross-Pitaevskii equation from the 
Bogoliubov-de Gennes equations. However, because of their approximate scheme, they end up with
the Born approximation $2a$ for the dimer scattering length $a_M$ instead of the exact result. 

In this paper, we present an exact fermionic theory of a BEC superfluid of composite bosons in the low density range. Our framework is completely general. Our present work is a first step toward going to higher orders, which will be clearly more complex to handle. Here we restrict ourselves to the $T=0$ thermodynamics. We obtain for the expansion of the chemical potential $\mu$ of our fermions of single spin density $n$ in the BEC regime:
\bea \label{eqlhy}
\mu &=&-\frac{E_b}{2}+\frac{\pi a_M}{m}\,n\,\left[\,1+\frac{32}{3\sqrt{\pi}}(na_M^3)^{1/2}\right]
\eea
Except for the obvious first term (which implies $\mu <0$), this is exactly 
the result found, for $\mu _{Bose}=2\mu $, by Lee, Huang and Yang (LHY)  \cite{lhy} for elementary bosons with density $n$, mass $m_B=2m$ and scattering length $a_B=a_M$. The identity of the mean field term is naturally somewhat expected. However,
even if it is reasonable to expect in our case a corrrection of the LHY type, it is not at all obvious that the coefficient is the same. We will see that our derivation is quite involved and has no systematic mapping 
on a purely bosonic formulation. In other terms
one expects the composite nature of our bosons to enter at some stage in the theory. We find indeed that this happens, but only at the
level of the $n^2$ term in Eq.(\ref{eqlhy}). Hence we prove that, for our composite bosons, the LHY term itself is unchanged with respect to elementary bosons \cite{longpaper}.
 
In order to perform a low density expansion, we need a "small parameter" in our theory. The most convenient one turns out to be the anomalous self-energy $\Delta (k)$ which, together with the anomalous (or off-diagonal) Green's function $F(k)$, is the hallmark of the superfluid state in the diagrammatic technique \cite{AGD}. We will indeed see that at low density $\Delta (k)$ is of order $n^{1/2}$, which could be anticipated from the standard BCS calculation \cite{leggett,popov}. Hence by performing an expansion in powers of $\Delta(k)$ in Feynman diagrams, we actually perform a low density expansion. The full Green's function $G(k)$ and self-energies are related by the completely general standard equations:
\begin{eqnarray}\label{geneq}
G(k)&=&{\mathcal G}_0(k) +F(k) \Delta^{*} (k) {\mathcal G}_0(k) \\  \label{eqdysond}
F(k)&=&-G(k) \Delta(k) {\mathcal G}_0(-k)
\end{eqnarray}
where we have set $[{\mathcal G}_0(k)]^{-1}=G_0^{-1}(k)-\Sigma(k)$, with $\Sigma(k)$ the normal self-energy, $G_0^{-1}(k)=\omega -k^2/2m + \mu$ and $k \equiv \{{\bf k},\omega\}$.

\begin{figure}[h]
\begin{center}
\vspace{-2mm}
\hspace{-10mm}\rotatebox{270}{\includegraphics[width=7cm]{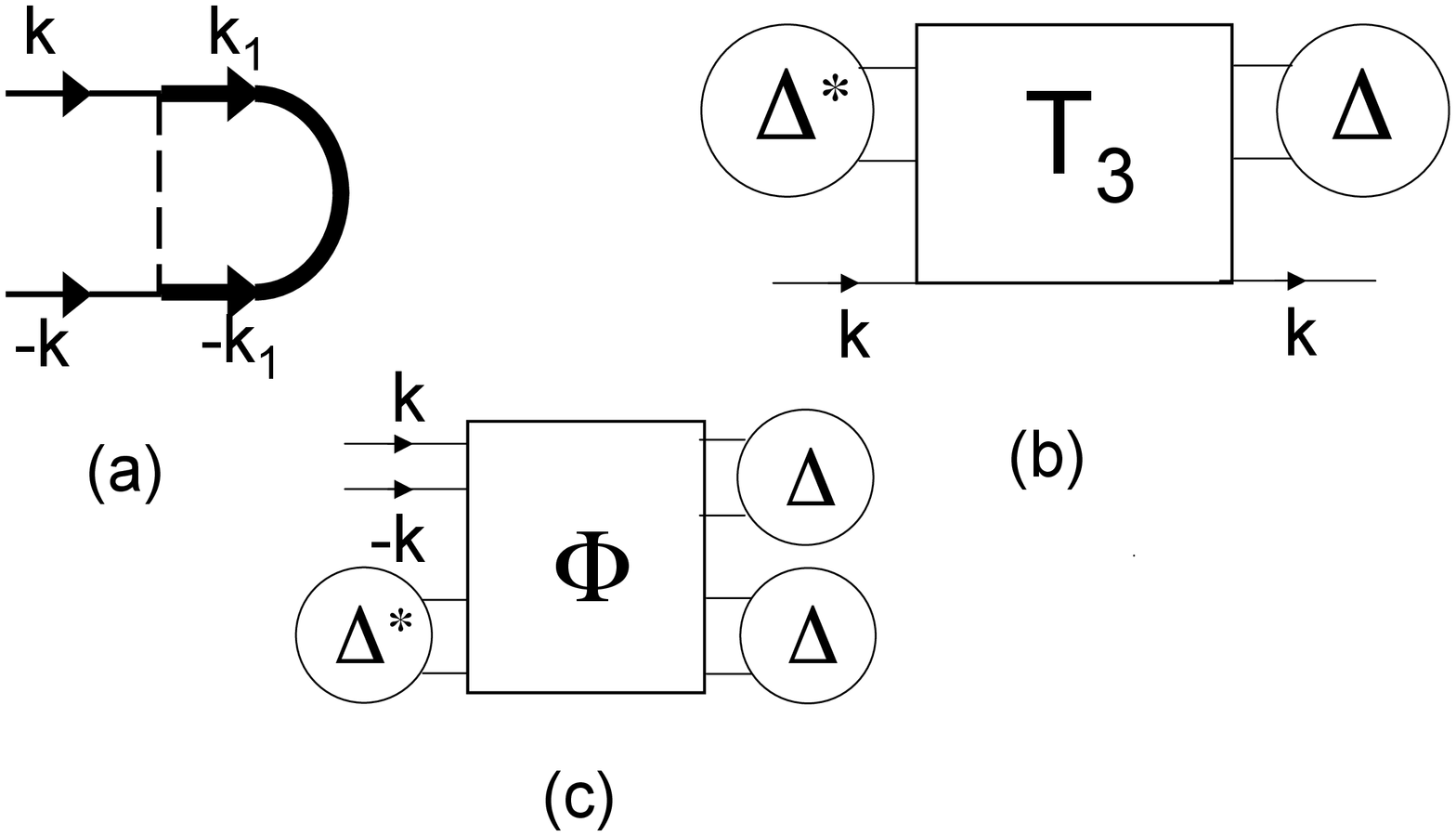}}
\vspace{-33mm}
\caption{(a) BCS-like contribution $\delta_1(k)$ (b) Structure of the lowest order normal self-energy (c) The diagram for $\delta_2(k)$. }
\label{figbcssig2F3}
\vspace{-8mm}
\end{center}
\end{figure}

We proceed in a natural way by finding the expansion of the Green's function $G$ and $F$ in powers of $\Delta$ at fixed $\mu $. The single spin density gives the "number equation":
\bea \label{numbeq}
n&=&-\sum_k e^{i\omega 0_+} G(k)
\eea
with $\sum_k \equiv i \int \frac{d^3 {\bf k}}{(2\pi)^3}
\int_{-\infty}^{+\infty}\frac{d\omega}{2\pi}$. At zeroth order in $\Delta (k)$ the result is obviously $n=0$ since, without condensate, there are no fermions at $T=0,\mu <0$. From particle conservation, the lowest order is given by the second order term:
\begin{eqnarray}\label{n2ord}
n_2=-|\Delta |^2\,\sum_k\; e^{i\omega 0_+} T_3(k,k;k)\, \left[G_0(k)\right]^2
\end{eqnarray}
where $T_3$, depicted in Fig.\ref{figbcssig2F3}(b), has been discussed in Ref.\cite{condmat4par,gurarie} and
contains all the normal state diagrams describing the scattering of a single atom by a dimer (actually in the involved vacuum Green's functions we have to shift the frequencies by the chemical potential $\mu$). This includes in particular a term $-G_0(-k)$ which is just the Born approximation for $T_3$. In writing Eq.(\ref{n2ord}) we have made use of the fact that, at this order, the $k$ dependence of $\Delta (k)$ can be neglected as will be shown below, and we have just denoted the resulting constant by $\Delta $. The frequency integral in Eq.(\ref{n2ord}) can be calculated by closing the contour in the upper-half complex plane $\mathrm {Im}\, \omega >0$, where $G_0(k)$ is analytic. It can be proved that, except for the Born term, $T_3(k,k;k)$ is also analytic in this half-plane. Hence the only contribution in Eq.(\ref{n2ord}) comes from the Born term. However this term is the only one considered in the standard BCS theory on this BEC side. We end up with the very surprising conclusion that, at this order, all the detailed physics involved in the atom-dimer scattering is irrelevant and that the result is merely given by the standard BCS calculation, namely $n_2=m^2\,|\Delta |^2/[8\pi\,(2m|\mu |]^{1/2})$. This shows that $\Delta$ is indeed of order $n^{1/2}$.

We consider now the anomalous self-energy $\Delta(k)$, in order to obtain our equivalent of the "gap equation" \cite{longpaper}.
$\Delta(k)$ describes two atoms $(k \uparrow, -k \downarrow )$ which go in the condensate. Quite generally the contributions to $\Delta(k)$ are divided in two classes, so we have $\Delta(k)=\delta_1(k)+\delta_2(k)$. The first class, the only one found in BCS theory, gathers diagrams where these two fermions \emph{first} interact through the bare two-body potential, with Fourier transform $V({\bf q})$, the second class containing all the other possibilities. In  full generality the contribution of the first class, shown diagrammatically in Fig.\ref{figbcssig2F3}(a), can be written: 
\bea
\delta_1({\bf k})&=&-\sum_{k_1}V({\bf k}-{\bf k_1})F(k_1)
\label{eqd1}
\eea
from the very definition of the {\it full} Green's function $F$.
Note that $\delta_1({\bf k})$ is independent of $\omega$ and, since the potential is very short-ranged, it depends on ${\bf k}$ only for very high momenta.

In the second class, where the two incoming
 fermions $(k \uparrow, -k \downarrow )$ do {\it not first} interact, we proceed to a $\Delta $ expansion. The first order term is already included in $\delta_1$ and particle conservation implies that the next order contains 
$\Delta \Delta^{*} \Delta $, the rest of the diagrams being made only of any number of normal state propagators $G_0(k)$ with any number of interactions, as shown schematically in Fig.\ref{figbcssig2F3}(c). Moreover these diagrams can not contain loops of normal state propagators, since
this would correspond, in time representation, to the creation of particle-hole pairs. Such processes are impossible in the normal state at $T=0$ and $\mu <0$, where the free particle propagator is {\it retarded}.
When these constraints are taken into account, including the 'no first interaction' condition, one ends 
up with the conclusion that these normal state diagrams have exactly been considered in Ref.\cite{condmat4par} (with again a trivial shift of all the frequencies by $\mu $, as for $G_0(k)$), and denoted by $\Phi$, except for a subtle point which we discuss below and is accounted for by the slightly different notation $\Phi'$. Hence:
\begin{eqnarray}\label{eqdelta2}
\delta_2(k)=\frac{1}{2}\, |\Delta|^2 \Delta\, \Phi'(k,-k;0,0)
\end{eqnarray}
In writing Eq.(\ref{eqdelta2}) we have taken advantage that, to lowest order (see Eq.(\ref{eqd1})), $\Delta (k)$ is a constant independent of $k$. Hence in this third order term, we can take $\Delta (k)$ as constant. On the other hand it is clear from Eq.(\ref{eqdelta2}) itself that $\Delta (k)$ depends in general on $k$. The factor $1/2$ is required to avoid double counting which arises from the presence of two factors $\Delta $.

The difference between $\Phi$ and $\Phi '$ stems from the fact that $\Phi$ is reducible, while $\Phi '$ is not since it is
a contribution to the anomalous self-energy. Specifically $\Phi$ contains the contribution $-G_0(k)G_0(-k)$ (this is the Born term) and also a term arising from the normal self-energy $\Sigma(k)$.
Hence, in order to obtain $\Phi '$, one has to subtract from $\Phi$ these reducible diagrams. However exactly these same reducible diagrams appear automatically if we write from Eq.(\ref{geneq},\ref{eqdysond}) the series expansion for $G_0^{-1}(k)F(k)G_0^{-1}(-k)$ in terms of the (irreducible) self-energies $\Delta (k)$ and $\Sigma (k)$. Hence it is more convenient to add these reducible contributions on both sides of the equation for $\Delta (k)$, in which case we have $-G_0(k)^{-1}F(k)G_0^{-1}(-k)$ in the left-hand side and $\Phi$ appears in the right-hand side, instead of $\Phi '$ (note that this manipulation is actually valid
to any order in our expansion). This leads to:
\begin{eqnarray}\label{eqf}
\hspace{-25mm}-F(k)=G_0(k)\delta_1({\bf k})G_0(-k) \nonumber
\end{eqnarray}
\vspace{-8mm}
\begin{eqnarray}\label{}
\hspace{15mm}+\frac{1}{2}\, |\Delta|^2 \Delta\, G_0(k)G_0(-k)\Phi(k,-k;0,0)
\end{eqnarray}
We then eliminate $F(k)$ in favor of $\delta_1$ by making use of Eq.(\ref{eqd1}). The summation of the last term over $k$ introduces \cite{condmat4par} the dimer-dimer scattering vertex $T_4(0,0;0)= \sum_{k}G_0(k)G_0(-k)\Phi(k,-k;0,0)$ evaluated at zero dimer energy. It is directly related \cite{condmat4par} to the dimer scattering length by $ (8\pi/am^2)^2T_4(0,0;0)=4\pi a_M/m$.
The last step in our procedure is the standard elimination of the interaction potential in favor of the scattering amplitude \cite{galit}. In our case this quantity has to be evaluated at the energy $\mu $, because our shift in frequency. After this step, $\delta_1({\bf k})$ can be taken as a constant $\delta_1$, since all the momentum integrals are rapidly convergent. We end up with:
\begin{eqnarray}\label{mean}
a^{-1}-\sqrt{2m|\mu|}&=&\frac{m^2 a^2}{8}\, a_M\,|\Delta|^2
\end{eqnarray}
where we have simplified by $\delta_1$ and made $\delta_1 \simeq \Delta $ in the right-hand side.
When we substitute for $|\Delta|^2$ its lowest order expression found above in terms of $n_2$, we find for $\mu $ the mean field part of Eq.(\ref{eqlhy}), with the appropriate dimer scattering length $a_M$.

The above is only the first step in our derivation. The natural continuation would be to go to next order
in $\Delta $, i.e. to order $\Delta ^4$ in Eq.(\ref{numbeq}) and order $\Delta ^5$ in Eq.(\ref{eqf}). This would lead to a contribution of order $n^2$ in Eq.(\ref{eqlhy}). However this expansion is not regular, as it would be the case if we had a gap between the ground state and the first excited state. Indeed there is, in our neutral superfluid, a branch of the excitation spectrum which goes to zero energy when momentum is zero. This is the collective mode, physically identical to sound waves in the low energy range, which is known as the Bogoliubov mode for Bose-Einstein condensates of elementary bosons. Naturally its existence is a fundamental property of the condensate
\cite{CKS}. In the following we include only the contributions coming from this low energy collective mode. 

The propagator of this collective mode is a two-particle vertex and it is the generalization to the superfluid state of the normal state dimer propagator $T_2(P)$. It enters our formalism in the following way. In Fig.1 b) and c), the terms $\Delta ^{*}$ and $\Delta $ act as 'source' and 'sink' of fermions. They are required since, at $T=0,\mu <0$, no fermions are present except coming from the superfluid. However we can in general well think of having a dimer propagator going from $\Delta $ to $\Delta ^{*}$ (and replacing them) as shown in Fig.3. This plays the same role for 'source' and 'sink', and gives diagrams which must be considered. It is easy to see that, in the normal state, they give a zero contribution (since there are no dimers in the normal state). But in the superfluid state, this dimer propagator has to be replaced by the collective mode and the result is non zero. The terms we have to retain are just the modifications, with respect to the previous results, coming from this substitution. Actually we do not proceed immediately to a $\Delta $ expansion and our procedure is equivalent to series resummation to avoid singularities.

To proceed we have first to find in our framework the collective mode propagator, more specifically in the low energy, low momentum range. It has a normal part $\Gamma(P)$ and an anomalous part $\Gamma^a(P)$, which both depend only on the total energy-momentum $P \equiv \{{\bf P},\Omega\}$. We write for them the equivalent of Eq.(\ref{geneq},\ref{eqdysond}), i.e. the Bethe-Salpeter equations:
 \bea \label{eqgam}
 \Gamma&=&T_2+T_2\,\Gamma_{\mathrm irr}\,\Gamma
+T_2\,\Gamma^a_{\mathrm irr}\,\Gamma^a\\ \label{eqgama}
\Gamma^a&=&
T_{2-}\,\Gamma_{\mathrm irr -}\,\Gamma^a
+T_{2-}\,{\bar \Gamma}^a_{\mathrm irr}\,\Gamma
\eea
where we did not write explicitely the arguments which are $P$ or $-P$: for instance $T_2$ stands for $T_2(P)$ and $T_{2-}$ for $T_2(-P)$. The normal ($\Gamma_{\mathrm irr}$) and anomalous (
$\Gamma^a_{\mathrm irr}$ and ${\bar \Gamma}^a_{\mathrm irr}$) irreducible vertices are analogous to self-energies.
 
\begin{figure}[h]
 \vspace{-2mm}
\begin{center}
\vspace{-13mm}
\hspace{-15mm}\rotatebox{270}{\includegraphics[width=8cm]{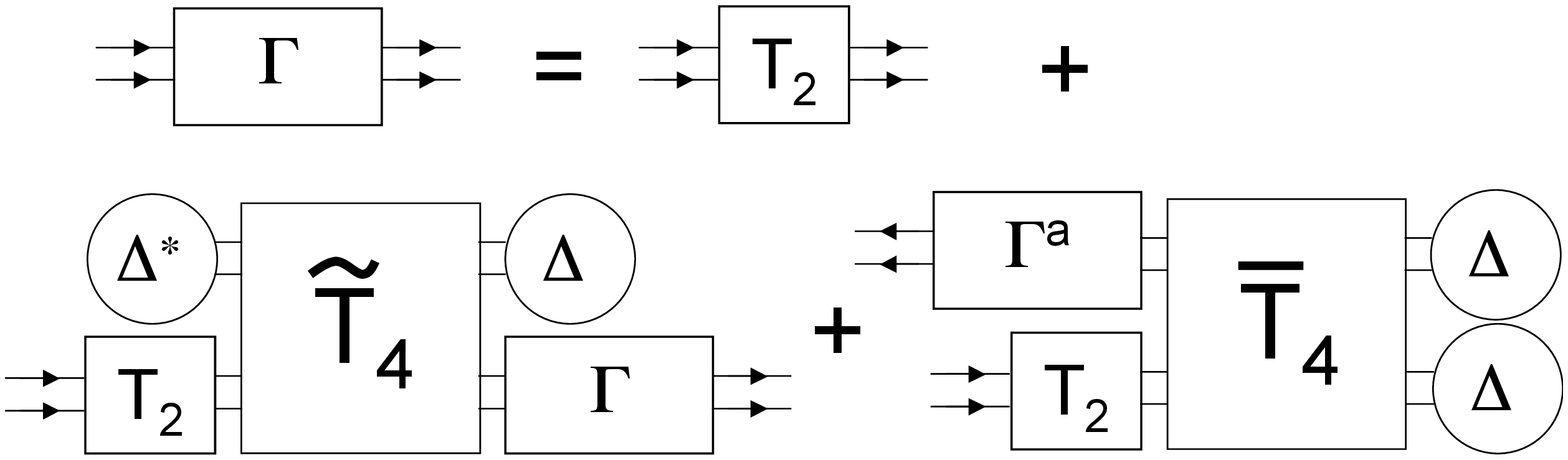}}
\vspace{-38mm}
\caption{Diagrammatic representation for Eq.(\ref{eqg11}) for $\Gamma$}
\label{figmodes11}
\end{center}
\end{figure}

Then we expand these irreducible vertices in powers of $\Delta$. Again from particle conservation the lowest order terms are second order. The result for Eq.(\ref{eqgam}) is depicted in Fig.\ref{figmodes11}. The 'normal part' (i.e. without the $\Delta$ factors) of the irreducible vertices involves clearly the normal state
dimer-dimer scattering vertex $T_4$ considered above, since all 'in' and 'out' lines are dimer lines. Again, at this lowest order, $\Delta$ can be taken as constant. In this way Eq.(\ref{eqgam}) and (\ref{eqgama}) become:
 \bea
 \Gamma&=&T_2+T_2\,|\Delta|^2 \,
\tilde{T}_4\,\Gamma\,+\,T_2\, \Delta^2\, \bar{T}_4\,\Gamma^a\label{eqg11}\\
\Gamma^a&=&T_{2-}\,|\Delta|^2\, \tilde{T}_{4-}\,\Gamma^a \,+\,T_{2-}\,\Delta^{*2}\, \hat{T_4}\,\Gamma\label{eqg12}
 \eea
 where $\tilde{T}_4= T_4(P/2,P/2;P/2)$, $\bar{T}_4=(1/2)T_4(P,0;0)$
 and $\hat{T_4}=(1/2)T_4(0,P;0)$, the factor $1/2$ being again topological.
\vspace{-6mm}
 
 We can now solve for $\Gamma$ and $\Gamma^a$. In the
 low energy limit $|{\bf P}| \ll 1/a$ and $|\Omega| \ll 1/ma^2$, we find easily 
 $\Gamma(P)=-8\pi (\Omega +\mu_B+{\bf P}^2 /4m)/(m^2aD)$ and $
\Gamma^a(P)=8\pi\mu_B/(m^2aD)$,
where $D=({\bf P}^2 /4m)^2 +2\mu_B\,  {\bf P}^2 /4m -\Omega^2$. We have set $\mu _B \equiv |\Delta|^2 ma\,a_M/4$
and evaluated the factor of $\Delta $ to zeroth order by taking $2|\mu |=1/ma^2$. The collective mode frequency is obtained by setting $D=0$ and we recover as expected the Bogoliubov dispersion relation.

\begin{figure}[h]
\begin{center}
\vspace{-3mm}
\hspace{-10mm}\rotatebox{270}{\includegraphics[width=6cm]{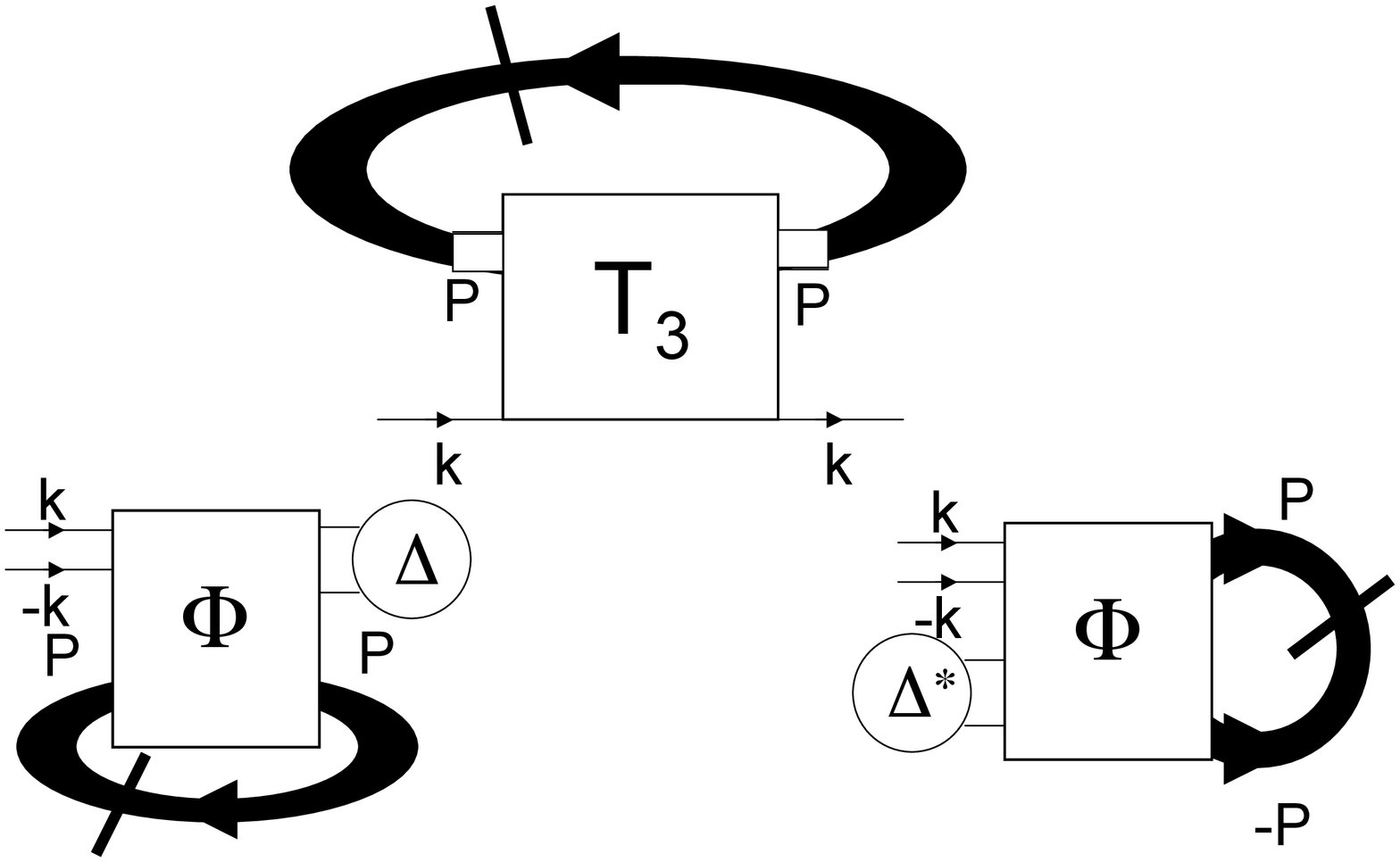}}
\vspace{-23mm}
\caption{Collective mode contributions to the self-energies}
\label{figsig3F4}
\end{center}
\end{figure}

We now consider the additional contributions to the self-energies coming from the collective mode. For the normal self-energy we have to add the top diagram in Fig.\ref{figsig3F4}, which gives an additional contribution $n_{cm}$ to our lowest order result Eq.(\ref{n2ord}):
\begin{eqnarray}\label{ncm}
n_{cm}=-\,\sum_{k,P}\; e^{i\omega 0_+} T_3(k,k;k+P)\, \Gamma(P) \left[G_0(k)\right]^2
\end{eqnarray}
Actually we should have subtracted from $\Gamma(P)$ its zeroth and second order terms in the series
expansion in powers of $\Delta $ (this is indicated in Fig.\ref{figsig3F4} by the slash in the mode propagator), since they are in principle taken into account in Eq.(\ref{n2ord}). However
it is easily seen that they are zero since they contain normal state propagator loops. In Eq.(\ref{ncm}) we can first perform the integration over the frequency variable $\omega $ of $k$, by closing the contour in the upper half-plane. Just as above in Eq.(\ref{numbeq}), it can be proved that the only contribution comes from the Born term of $T_3(k,k;k+P)$. Then the ${\bf k}$ integration is easily performed and we are left with $n_{cm}=(m^{3/2}/8\pi )\sum_P \Gamma(P )/[2|\mu |+{\bf P}^2/4m-\Omega ]^{1/2}$. The $\Omega $ integration can be transformed over a contour which encloses all the singularities of $\Gamma(P)$ on the real negative axis. The high energy contributions to $n_{cm}$ coming from $|\Omega| \gtrsim 1/ma^2$ (physically linked to breaking dimers) will give negligible regular terms of order $\Delta ^4$, as discussed above. On the other hand the contribution of the low frequency collective mode is easily calculated with the low energy expression of $\Gamma(P)$ given above. We find:
\begin{eqnarray}\label{eqcm}
n_{cm}= \frac{1}{3\pi^2}(2m\mu_B)^{3/2}
\end{eqnarray}
where we have used the zeroth order expression $E_b/2$ for $|\mu |$. When we use for $\mu_B$ its lowest order expression in terms of $n$, we find that $n_{cm}$ coincide formally with the 'depletion of the condensate', known for elementary boson  superfluids.
\vspace{-5mm}

We proceed now in the same way for the collective mode contributions to the anomalous self-energy. Corresponding to the diagram Fig.\ref{figbcssig2F3}c), we have to add the two bottom diagrams in Fig.\ref{figsig3F4}. Just as in Eq.(\ref{eqdelta2}) we should take only irreducible diagrams into account. But handling this problem in the same way by adding the reducible contributions on both sides of the equation, we end up with Eq.(\ref{eqf}) except, in the right-hand side, for the additional contribution $ \Delta\, G_0(k)G_0(-k)\Phi(k,-k;0,0)[ \sum_{P}\Gamma(P)+(1/2) \sum_{P}\Gamma^a(P) ]$. As in Eq.(\ref{eqdelta2}) the factor $1/2$ is topological. Then we follow the same procedure as after Eq.(\ref{eqf}). Finally, as in the calculation of $n_{cm}$, we retain only in the summation over $P$ the low energy contribution, the other ones giving higher order terms. The summation $ \sum_{P}\Gamma(P)$ has already been found
in the above calculation of $n_{cm}$. The summation $\sum_{P}\Gamma^a(P)$ is more involved since, as we mentionned earlier below Eq.(\ref{ncm}), we have to subtract from $\Gamma^a(P)$ the lower order terms already taken into account in our lowest order calculation, leading to Eq.(\ref{mean}). In contrast with the case of $\Gamma(P)$, the term we subtract is not zero, but acts to regularize the remaining integral over momentum ${\bf P}$, which would otherwise have a high momentum divergence \cite{rem}.
We obtain finally for the slashed contribution, which takes into account this subtraction, $\sum_{P}{\not \!\Gamma}^a(P)=3 \sum_{P}\Gamma(P)=
24\pi n_{cm} (2m|\mu |)^{1/2}/m^2$  

If we gather all the contributions, we have for the single spin density $n=n_2 + n_{cm}$ while the "gap equation" Eq.(\ref{mean}) is changed into the simple form:
\begin{eqnarray}
a^{-1}-\sqrt{2m|\mu|}&=&\frac{m^2 a^2}{8}\, a_M\,|\Delta|^2 +5\pi a\, a_M\,n_{cm}
\end{eqnarray}

When $|\Delta|^2$ is eliminated between the "gap" and the "number" equations, the consistently
expanded result for $\mu $ is indeed found to be Eq.(\ref{eqlhy}).

In conclusion we have shown how an exact purely fermionic framework can be used in the BEC regime
of the BEC-BCS crossover, and we have specifically demonstrated that the Lee-Huang-Yang result for the chemical potential remains valid for the corresponding composite bosons. We are very grateful to M. Yu. Kagan for stimulating discussions at the beginning of this work.


\begin{thebibliography}{99}
\bibitem{leggett}A. J. Leggett, J. Phys. (Paris), Colloq. {\bf 41}, C7-19 (1980).
\bibitem{Grei} M. Greiner, C. Regal, and D. Jin, Nature (London) {\bf 426}, 537 (2003).
\bibitem{Joch} S. Jochim, M. Bartenstein, A. Altmeyer, G. Hendl, S. Riedl, C. Chin, J. H. Denschlag, and R. Grimm, Science {\bf 302}, 2101 (2003).
\bibitem{Zwie} M. W. Zwierlein, C. A. Stan, C. H. Schunck, S. M. F. Raupach, S. Gupta, Z. Hadzibabic, and W. Ketterle, Phys. Rev. Lett. {\bf 91}, 250401 (2003).
\bibitem{Bour} T. Bourdel, L. Khaykovich, J. Cubizolles, J. Zhang, F. Chevy, M. Teichmann, L. Tarruell, S. J. J. M. F. Kokkelmans, and C. Salomon, Phys. Rev. Lett. {\bf 93}, 050401 (2004).
\bibitem{PSS}D. S. Petrov, C. Salomon, and G. V. Shlyapnikov, Phys. Rev. Lett. {\bf 93},
090404 (2004).
\bibitem{condmat4par} I.V. Brodsky, M. Yu. Kagan, A.V. Klaptsov, R. Combescot
and X. Leyronas, Phys. Rev. A {\bf 73}, 032724 (2006).
\bibitem{popov} V. N. Popov, Zh. Eksp. Teor. Phys. {\bf 50}, 1550 (1966),
[Sov. Phys. JETP {\bf 23}, 1034 (1966)].
\bibitem{kk}L. V. Keldysh and A. N. Kozlov, Zh. Eksp. Teor. Phys. {\bf 54}, 978 (1968)
[Sov. Phys. JETP {\bf 27}, 521 (1968)].
\bibitem{pieristrinati} P. Pieri and G. C. Strinati, Phys. Rev. Lett. {\bf 91}, 030401 (2003) and references therein for earlier work.
\bibitem{lhy} T. D. Lee and C. N. Yang, Phys. Rev. {\bf 105}, 1119 (1957);
T. D. Lee, K. Huang and C. N. Yang, Phys. Rev. {\bf 106}, 1135 (1957)
\bibitem{longpaper}Here we sketch the derivations. Details will be given in an extended version, in preparation.
\bibitem{AGD}  A. A. Abrikosov, L. P. Gorkov and I. E. Dzyaloshinski,
{\it Methods of quantum field theory in statistical physics} (Dover, 1975).
\bibitem{gurarie}J. Levinsen, V. Gurarie, Phys. Rev. A {\bf 73}, 053607 (2006).
\bibitem{galit}V.M. Galitskii, Sov. Phys. JETP, {\bf 7}, 104(1958).
\bibitem{CKS}See R. Combescot, M. Yu. Kagan, and S. Stringari, Phys. Rev. A {\bf 74}, 042717 (2006) for a study of this mode throughout the BEC-BCS crossover, within the (approximate) dynamical BCS theory (which gives $a_M=2a$).
\bibitem{rem}This subtraction removes also an (imaginary) contribution arising from $T_4(0,0;0)$, due to the fact that it has to be evaluated at the chemical potential $\mu$ which is not exactly half the bound state energy of the molecule. A similar feature is present in the approach of S. T. Beliaev, Sov. Phys. JETP {\bf 7}, 289 (1958) for elementary bosons.
\end{thebibliography}
 \end{document}